# Dark matter an effect of gravitation permeability of material in Jordan, Brance – Dicke theory.

V.I.BASHKOV, S.M.KOZYREV

*Department of physic, Kazan State University*
e-mail: Sergeym@space.com

## ABSTRACT

Analyzing the static spherically symmetric and rotating ellipsoid solutions in the Newtonian limit of Jordan, Brance – Dicke theory we find the following. In empty space scalar-tensor theories have trivial solution of field equation with constant scalar potential (efficient value of gravitation constant). In this case no celestial-mechanical experiments to reveal a difference between scalar-tensor theories and Einstein theory is not presented possible. However, scalar field, inside the matter, has characteristics like gravitation permeability of material similar electromagnetic permeability of material in Maxwell theories of electromagnetism. Investigation of obtained exact solutions for given functions of a matter distributions in the Newtonian limit of Jordan, Brance – Dicke theory show the efficient value of gravitation constant depends on density of matter, sizes and form of object, as well as on the value of theories coupling constant. That for example led to weakening gravitation force in the central regions of a Galaxies. This assumption constitutes the way to explain observed rotation curves of Galaxies without using cold dark matter.

## 1. INTRODUCTION

Scalar-tensor theories are among the most viable alternative to Einstein's general theory of relativity. There are the mathematical difficulties of these theory lies in the high non-linearity of the field equations. However, when the gravitational field is weak one can linearize the field equations. In these approximations, all these theories should be expected to reproduce the Einstein general relativity, which is experimentally tested only in this limit. The method for approximate solutions of field equations of different gravitation theories for low-energy particles systems, has been found by Eddington and Robertson [1,2]. This method are interconnected by gravitation forces, is post-Newton approximation to decomposition of metric coefficient for spherically symmetric static case into degree ranges according to a small parameter $1/c^2$.

As shown in [3], solar-system experiments allowed one to map out fairly completely weak-field gravity at the first post-Newtonian approximation, i.e., to put stringent numerical constraints on a large class of possible deviations from general relativity at order $1/c^2$. However, post-Newtonian parameters of the concrete gravitational theory under study will take different values in external and internal areas of material objects [4,5]. Moreover, in empty space scalar-tensor theories have trivial, at first thought, solution of field equation: $f$ = const. In this case no celestial-mechanical experiments to reveal a difference between scalar-tensor theories (as well as vector-metric theories) and Einstein theory is not presented possible, since all Einstein's vacuum solution (Schwarzschild, Kerr, etc.) will satisfy this theories too. Having taken the case of this solution we come to a conclusion that PPN parameters of external solution in scalar-tensor theories and general relativity are similar.

In this paper, we want to discuss the Newtonian limit of Jordan, Brance - Dicke (JBD), the simplest of the scalar-tensor theory of gravity. In particular, we show that in Newtonian limit of JBD scalar field, inside the matter, have characteristics like gravitation permeability of material.

In Sec. 2 and Sec. 3 we discuss the general assumptions made in this paper. In Sec. 4, we obtain static, spherical and ellipsoidal symmetric solutions for a few functions of a matter distributions. Structures of gravitation field discuss in this section. Analysis relation between $\phi$ and U made in Sec. 5. The rotation curves of galaxies are deduced in Sec. 6. JBD models constitute the way to explain rotation curves without using cold dark matter.

## 2. ACTION AND FIELD EQUATIONS

As it has been stated before, the most general theory containing a spin-2 field and one spin-0 field contains arbitrary coupling functions $\omega(\phi)$ and $\lambda(\phi)$. The theory can be expressed in units in which the local value of the Newtonian "gravitation constant" is a function of a scalar field which is turn determined by the trace of energy-momentum tensor of all other nongravitational fields. The action is given by

$$S = \int d^4 x \sqrt{-g} \, [16\pi L + \phi R - \omega(\phi) g^{mn} \frac{\partial_m \phi \partial_n \phi}{\phi} - \lambda(\phi)], \quad (1)$$

where $\lambda(\phi)$, $\omega(\phi)$ are generic functions of $\phi$. In equation (1) we have generalized the particle Lagrangian to be a matter Lagrangian which could be include electromagnetic fields and forces, nuclear forces etc. By varying S, one obtains

$$R_{mn} - \frac{1}{2} g_{mn} R = 8\pi \phi^{-1} T_{mn} + \phi^{-2} \omega(\phi) \left( \phi_{,m} \phi_{,n} - \frac{1}{2} g_{mn} \phi_{,l} \phi^{,l} \right) + \phi^{-1} (\phi_{,mn} - g_{mn} \Delta_g \phi) + \lambda(\phi) g_{mn},$$

$$\Delta_g \phi + \frac{1}{2} \phi_{,m} \phi^{,m} \frac{d}{d\phi} \ln\left( \frac{\omega(\phi)}{\phi} \right) + \frac{1}{2} \frac{\phi}{\omega(\phi)} \left[ R + 2 \frac{d}{d\phi} (\phi \lambda(\phi)) \right] = 0,$$

(2)

where $T_{\mu\nu}$ energy – momentum tensor of matter, $\phi$ scalar field, which is interpret as inverse "gravitation constant":

$$\phi \approx \frac{1}{G}$$

where G gravitation constant equal 1. We are used unit system c = G = 1. Equations (2) become identical with the JBD field equations when $\omega(\phi)$ is set to a constant $\omega$ and $\lambda(\phi)$ is equal to zero.

Vacuum solutions of equations (2) with $\lambda(\phi) = 0$, $\omega(\phi) = \omega = constant$, under assumptions of static spherically symmetric with metric in the form

$$ds^2 = -e^{2\gamma}(r)dt^2 + e^{2\lambda}(r)dr^2 + r^2(d\theta^2 + \sin^2\theta d\varphi^2),$$

was found by Heckmannn [6]. Obviously, for general case, spherically symmetric static solutions of this equation (2) give in empty space amongst the others:

$$\phi = 1$$
$$g_{00} = -g_{11}^{-1} = 1 - 2M/r$$
(3)

that is equivalent to Schwarzachild solution in general relativity. Thus, in this case all celestial mechanic experiments cannot reveal difference between these theories. Analogous result we can find for vector - metric theory

$$R_{mn} - \frac{1}{2} g_{mn} R + w\overset{(w)}{q}_{mn} + h\overset{(h)}{q}_{mn} + e\overset{(e)}{q}_{mn} + t\overset{(t)}{q}_{mn} = 8\pi G T_{mn}$$

$$eF^{mn}_{;n} + \frac{1}{2} tK^{mn}_{;n} - \frac{1}{2} wK^m R - \frac{1}{2} hK^n R^m_n$$

(4)

External solution for the case of spherically symmetric static distribution of matter may be:

$$g_{00} = -g_{11}^{-1} = 1 - 2M/r$$

$$\vec{K} = const$$

However, for the internal solutions increases the interaction between scalar, vector and tensor fields describing gravitation. As a consequence of that there is a significant difference in characteristics of JBD and Einstein's galaxy models under similar boundary conditions. So, in case of our suggestion is correct, effects considered like the fifth force and depends on the differences between JBD and an Einstein's object should be experimentally tested in substance namely.

### 3. WEAK FIELD APPROXIMATION.

In this section we discuss the weak-field limit of Jordan, Brans-Dicke theory [6, 7]. The JBD theory incorporates the Mach principle, which states that the phenomenon of inertia must arise from accelerations with respect to the general mass distribution of the universe. Differences between predictions of JBD theory and observing appeared at study of solving the field equation (2) where $\lambda(\phi) = 0$, $\omega(\phi) = \omega = constant$.

$$\Box \phi = \frac{8\pi}{3 + 2\omega} T^\mu_\mu \quad (5)$$

$$R_{\mu\nu} - \frac{1}{2} g_{\mu\nu} R = -\frac{8\pi}{\phi} T_{\mu\nu} - \frac{\omega}{\phi^2}\left(\phi_{;\mu}\phi_{;\nu} - \frac{1}{2} g_{\mu\nu}\phi_{;\rho}\phi^{;\rho}\right) - \frac{1}{\phi}(\phi_{;\mu;\nu} - g_{\mu\nu}\Box\phi) \quad (6)$$

It is well known that this theory is self-consistent, complete and for $|\omega| \geq 500$ in accord with solar system observations and experiments [3,8,9]. However, there are the solution (3) of equations (5,6) [4,5], in this case of external solution in JBD and general relativity are similar, and parameter $\omega$ is free.

The problem of finding internal analytical solutions of equations (5)-(6) are difficult, so qualitative analysis is realise by use weak field approximation. Equations of gravitation field allow significant simplification, if velocity of material point far less then velocity of light, so values of order $v^2/c^2$ possible neglect. In the case of weak field approximation value $g_{\mu\nu}$ must extremely little differ from:

$g_{\mu\nu} = 1$ for $\mu = \nu = 1, 2, 3$, $g_{00} = -1$

$g_{\mu\nu} = 0$ for $\mu \neq \nu$

and squares of these deflections possible to neglect. Then

$$\frac{d^2 x^m}{dt^2} = -c^2 \Gamma^m_{00} \quad (7)$$

Moreover, in static case derived $g_{\mu\nu}$ on time possible to neglect too. Then one can change $\Gamma^m_{00}$ to $\Gamma_{m,00}$, or $-\frac{1}{2}\frac{dg_{00}}{dx^m}$ and equations of motion a material point (7) for small velocities and weak field takes Newtonian form:

$$\frac{d^2 x^m}{dt^2} = -\frac{\partial U}{\partial x^m}$$

where U is a gravitation potential, and

$$g_{00} = 1 - \frac{2U}{c^2}. \qquad (8)$$

From equations (5)-(6) one can get

$$R_{mn} + \frac{w}{f^2}f_{;m}f_{;n} + \frac{1}{f}f_{;m,n} = -\frac{8p}{f}[T_{mn} - \frac{1+w}{3+2v}Tg_{mn}] \qquad (9)$$

For the component 00 equations (9) values $T_{mn}$ of order v/c possible neglect, except $T_{00}$. $T_{00} = \rho c^2$ consequently $T = g^{mn}T_{mn} = g^{00}T_{00} = -\rho c^2$ and

$$R_{00} + \frac{1}{f}f_{;0;0} = -\frac{1}{2f}k\rho c^2$$

As far as derived on time and product $\Gamma^m_{ns}$ we neglect, then

$$R_{00} = \frac{\partial \Gamma^i_{00}}{\partial x^i}$$

since $\Gamma^i_{00} \approx \Gamma_{i,00} \approx -\frac{1}{2}\frac{\partial g_{00}}{\partial x^i}$ then from (8)

$$R_{00} = \frac{1}{2}\sum_i \frac{\partial^2 g_{00}}{\partial x^2_i} = \frac{1}{2}\Delta g_{00} = -\frac{\Delta U}{c^2}$$

For scalar potential we have $f_{;0;0} = 2\Gamma^i_{00}\frac{\partial f}{\partial x^i}$.

Finely

$$div(f\nabla U) = 8p\frac{2+w}{3+2w}\rho \qquad (10)$$

$$\Delta f = -\frac{8p\rho}{3+2w} \qquad (11)$$

where ρ is density of mater. Limiting transformation to Newton theory of gravitation occurs when $|v| \to \infty$ and $f = constant$.

## 4. AN EXACT SOLUTIONS

For the further specification of the problem an equation of state is needed. As alternatives one can use functions of a matter distributions for spherically symmetric objects of radius R. For example there are the model of gas density distribution for clusters of galaxies [10], which in most cases provides an adequate description of the mean of X-ray intensity with projected radius:

$$\rho = \rho_0[1 + (\frac{r}{R})^2]^{-1.5b} \qquad (12)$$

where the core radius, R, central density $\rho_0$, and the number $b$ are parameters varies from cluster to cluster but has typical value the order of 2/3 which implies that the gas mass generally increases linearly with radius. Analogous form of density distribution there are for globular star cluster [11], in this case $b = 1$. In this section for qualitative analysis we use a few simple models.

$$r = r_0 \tag{13}$$

$$r = r_0 \left(\frac{R}{r} - 1\right) \tag{14}$$

$$r = r_0 (1 - e^{r-R}) \tag{15}$$

$$r = r_0 (R - r) \tag{16}$$

In empty space a solution (3) for Newtonian approximations of JBD theories stays correct. In this case outside the objects will be usual Newtonian universe, but inside "gravitation constant" will depend on matter distributions. Solution an equation (3) for considered distributions a material accordingly will:

$$f = \frac{1}{r(3+2w)} [4\sqrt{2}pR^2(R-r)r + r(3+2w) - 8pr(R^3 Log[R(1+\sqrt{2})] - \frac{R^4 \sqrt{1 + (\frac{r}{R})^2} Log(r + \sqrt{r^2 + R^2})}{\sqrt{r^2 + R^2}})] \tag{17}$$

$$f = 1 - \frac{4pr(r-R)^2(r+2R)}{3r(3+2w)} \tag{18}$$

$$f = 1 + \frac{4pr(r-R)^3}{3r(3+2w)} \tag{19}$$

$$f = \frac{1}{3r(3+2w)} \{r[9 + 12pr(2 + e^{r-R} + (R-2)R) + 6v] - 4pr^3 r + 8pr[6 - 6e^{r-R} - R(6 + (R-3)R)]\} \tag{20}$$

$$f = 1 + \frac{2pr(r-R)^3(r+R)}{3r(3+2w)} \tag{21}$$

From brought solutions (17)-(21) follows that scalar field shows characteristics like gravitation permeability of material. The efficient value of gravitation constant depends on density of matter and sizes of object, as well as from the value ω. As can be seen from (17)-(21) for any chosen functions of a matter distributions gravitation interaction of inwardly objects goes to zero in the despised centre and increases beside surfaces. When increasing a radius or density of configurations the gravitation interaction inwardly objects greatly weakens and it were " squeeze out " to the surface. Under the endless radius of configurations a value gravitation constant in all space becomes zero. As illustrations of stated conclusions will bring several graphs.

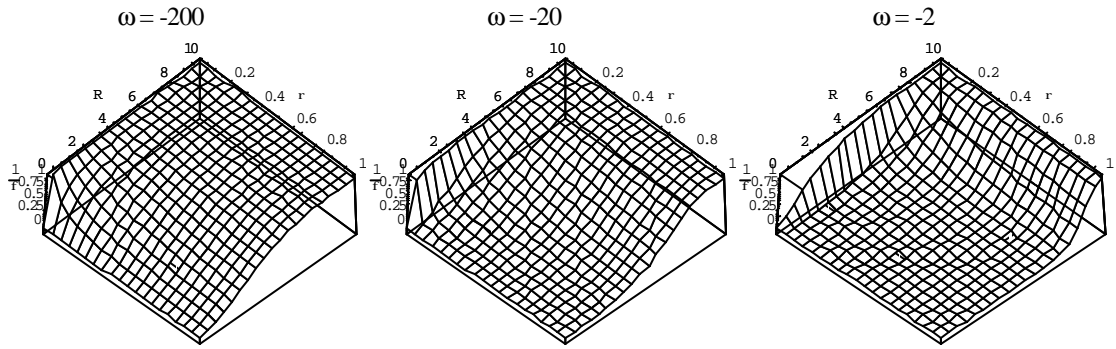

*Fig. 1. The gravitation "constant" for objects with $\rho = \rho_0 (R/r-1)$. For any chosen functions of matter distribution with increasing a radius or density of configurations gravitation interaction inwardly objects goes to zero in the despised centre and increases beside surfaces.*

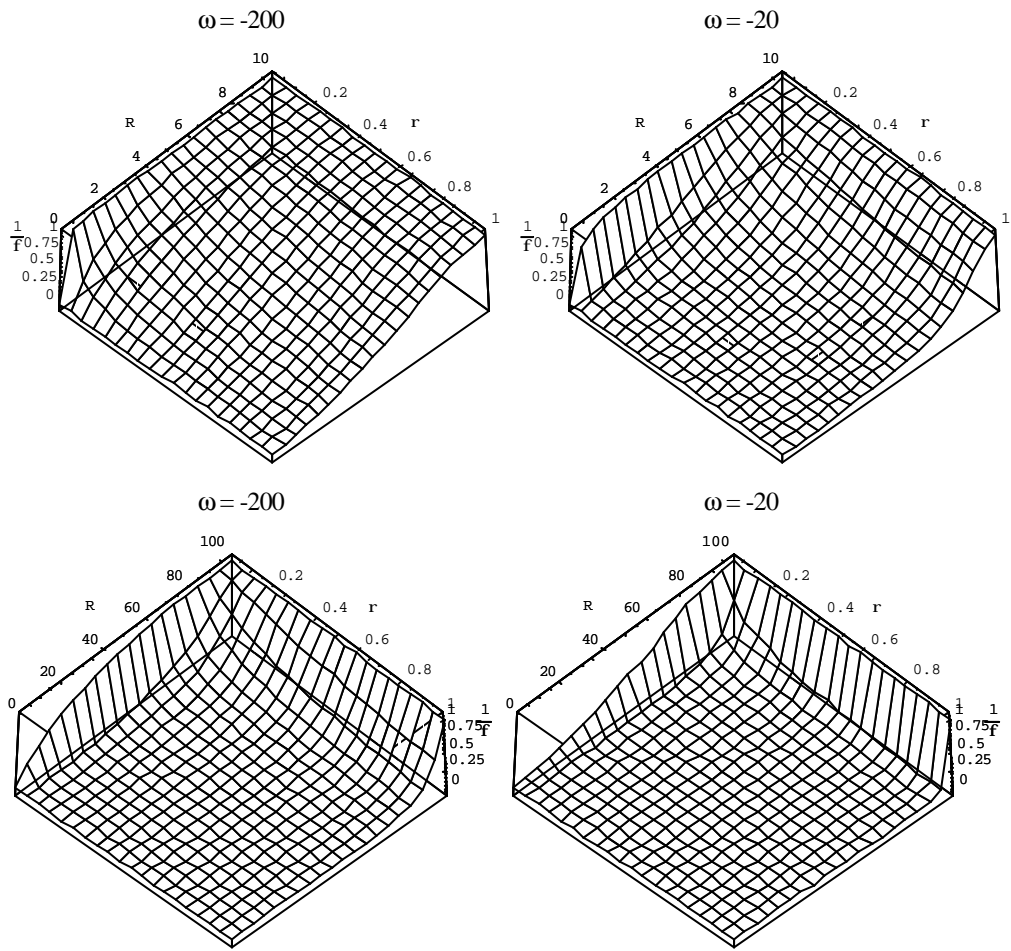

*Fig. 2. The gravitation "constant" for objects with constant density. When increasing a radius or density of configurations the gravitation interaction inwardly objects weaken.*

Solutions an equation (11) for objects having form of ellipsoids of rotating to manage one can find using spheroidal coordinates $x, h, j$ received by rotating the confocal ellipses around axis of symmetry. Equation for scalar field (11) in curvilinear coordinates has of the form of:

$$\frac{1}{h_1 h_2 h_3}\left[\frac{\partial}{\partial x}\left(\frac{h_2 h_3}{h_1}\frac{\partial f}{\partial x}\right) + \frac{\partial}{\partial \eta}\left(\frac{h_3 h_1}{h_2}\frac{\partial f}{\partial \eta}\right) + \frac{\partial}{\partial \varphi}\left(\frac{h_1 h_2}{h_3}\frac{\partial f}{\partial \varphi}\right)\right] = -\frac{8\pi\rho}{3+2\omega} \tag{22}$$

where $h_i$ is a coefficients Lame. For extended spheroids:

$$h_1 = q\sqrt{\frac{x^2 - h^2}{x^2 - 1}}$$
$$h_2 = q\sqrt{\frac{x^2 - h^2}{1 - h^2}} \tag{23}$$
$$h_3 = q\sqrt{(x^2-1)(1-h^2)}$$

and for compressed spheroids,

$$h_1 = q\sqrt{\frac{x^2 - h^2}{x^2 - 1}}$$
$$h_2 = q\sqrt{\frac{x^2 - h^2}{1 - h^2}} \tag{24}$$
$$h_3 = q\,xh$$

Solutions an equation (22) for a matter distributions:

$$\rho = \frac{\rho_0}{x^2 - h^2} \tag{25}$$

will accordingly:

$$f = 1 + \frac{4\pi\rho q^2\left((1-S)Log\frac{(x-1)}{(S-1)} + (1+S)Log\frac{(1+x)}{(1+S)}\right)}{3+2\omega}$$

$$f = 1 + \frac{8\pi\rho q^2\sqrt{S^2-1}(ArcCot(\sqrt{S^2-1}) - ArcCot(\sqrt{x^2-1})) + LogS - Logx}{3+2\omega}$$

where **S** border of configurations, **q** half of distance between focuses of ellipse.

Either as for the spherical symmetry, efficient value of gravitation constant depends on density of matter, sizes of object, values ω and, moreover, from the form of objects. Gravitation interaction inwardly rotating ellipsoids, as extended, so and compressed goes to zero in the centre of configurations and increases beside surfaces. When increasing a distance between focuses of configurations gravitation interaction weakens and " squeeze out " to the surface.

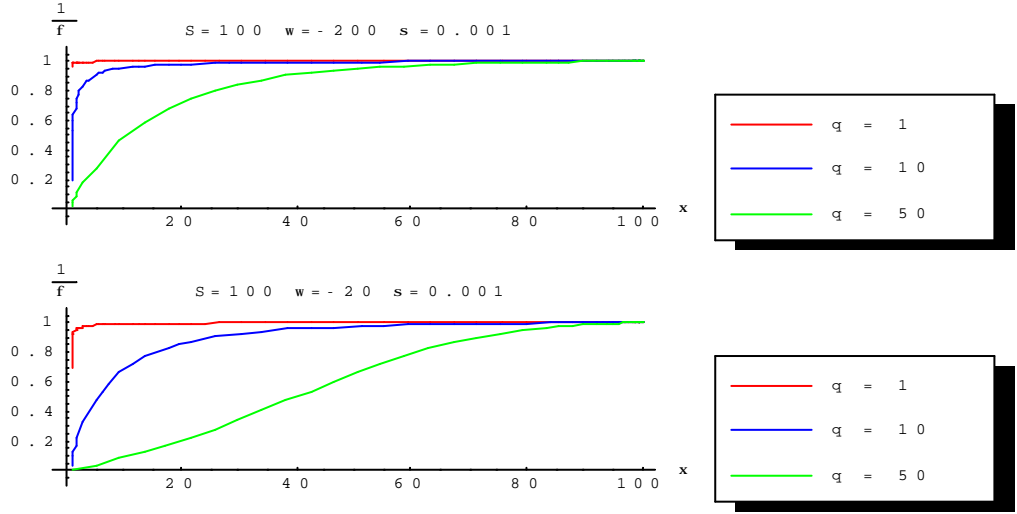

*Fig. 4. Gravitation constant for extended spheroids objects with $\rho = \rho_0 / (\xi^2-\eta^2)$. When increasing a distance between focuses of configurations gravitation interaction inwardly objects decreases.*

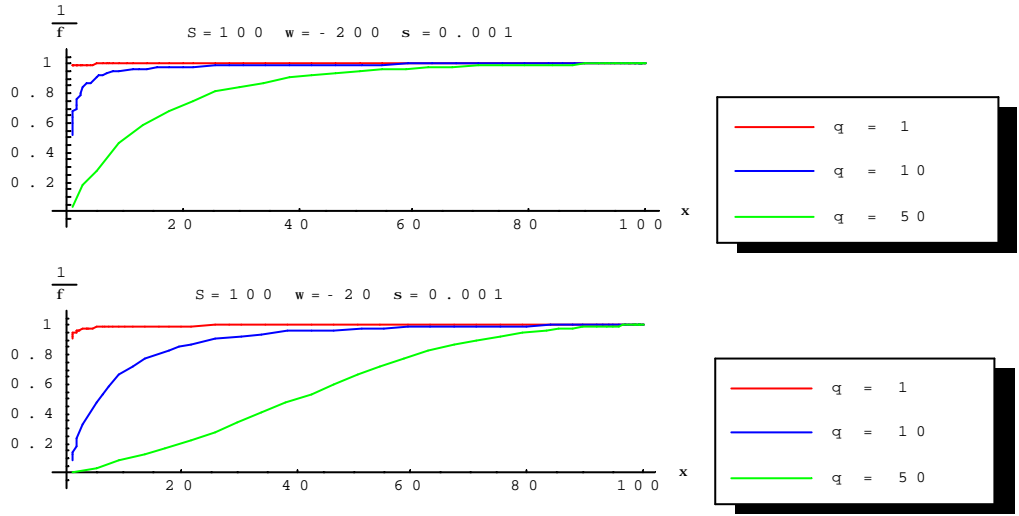

*Fig. 5. Gravitation constant for compressed spheroids objects with $\rho = \rho_0 / (\xi^2-\eta^2)$. When increasing a distance between focuses of configurations gravitation interaction inwardly objects decreases.*

## 5. ON A RELATION BETWEEN $f$ AND U.

For the perfect fluid model with simple equation of state, $p = \varepsilon\rho$, for matter there are an exact solution [12] of equations (5)-(6). When the spacetime is static and configuration is spherically symmetric the line element can be put in the Schwarzschild form

$$ds^2 = -e^{2\gamma}(r)dt^2 + e^{2\lambda}(r)dr^2 + r^2(d\theta^2 + \sin^2\theta d\varphi^2) \qquad (26)$$

The energy-momentum tensor $T_{\mu\nu}$ is specialised to that of a perfect fluid

$$T_{mn} = (p + \rho)V_m V_n + p g_{mn}, \qquad (27)$$

where $p$ and $\rho$ are proper pressure and energy density, respectively. $V_m$ are components of the fluid four-velocity. For the static fluid $T_0^0 = -\rho$; $T_1^1 = T_2^2 = T_3^3 = p$. From equation (6) under this condition we obtain

$$-\frac{1}{2}[\sqrt{-g} f (\ln g_{00})^{,k}]_{,k} = 8\pi(T_0^0 - \frac{1+\omega}{3+2\omega}T)\sqrt{-g}. \qquad (28)$$

Furthermore in this case equation (5) simplifies to

$$[\sqrt{-g}f(\ln f)^{,k}]_{,k} = \frac{8pT\sqrt{-g}}{3+2w}. \tag{29}$$

Combining equations (28)-(29) and making in use of equation of state, $p = er$, we get

$$\frac{s}{2}[\sqrt{-g}f(\ln g_{00})^{,k}]_{,k} - [\sqrt{-g}f(\ln f)^{,k}]_{,k} = 0 \tag{30}$$

or

$$(\sqrt{-g}f\{\ln[fg_{00}^{-\frac{c}{2}}]\}^{,k})_{,k} = 0, \tag{31}$$

where $s \equiv \dfrac{3e-1}{(3+2w)+(1+w)(3e-1)}$.

The similar result one can find in Newtonian approximation for any type equations of state. The form analogous to (31) of the field equations is derived from (10)-(11) by means of the transformation $u = e^U$,

$$div(f\nabla \ln \frac{f}{u^b}) = 0, \tag{32}$$

where $b = -\dfrac{8p}{2+w}$.

Assuming that spacetime is spherically symmetric, and that the metric chosen in the Schwarzschild form (26), the very special solution of (31) becomes

$$f = const \times e^{sy}. \tag{33}$$

In the Newtonian approximation, in simplest case one can obtain from (32)

$$f = const \times e^{bU}. \tag{34}$$

However, there are the big class of solutions equations (31) and (32) when the relation's (33) and (34) is not true. In a general case we mast solve the equations of gravitation and scalar fields without using the relation's (33) and (34).

## 6. THE ROTATION CURVE OF GALAXIES AND SCALAR FIELD.

There exists clear evidence for mass discrepancies in extragalactic systems. Application of the usual Newtonian dynamical equations to the observed luminous mass does not predict the observed motions. This leads to the inference of dynamically dominant amounts of dark matter.

The rotation curves of galaxies derived from gaseous tracers (Hα and H I) provide the strongest tests of force-laws [13]. With only the assumption of circular motion, it is possible to directly equate the centripetal acceleration [14]

$$a_c = \frac{v^2(R)}{R} \tag{35}$$

with the gravitational acceleration

$$g = -\nabla U$$

determined from the Poisson equation

$$\Delta U = 4pGr.$$

In the case of JBD Newtonian approximation value of a gravitation force $\nabla U$ define by solving the equations (10)-(11) with an equation of state or function of matter distributions. It can then be deduced there are the analytical solutions for the circular velocity is given for the all considered functions of matter distribution (12)-(16). From this solutions follows by choice of values ω possible explain rotation curve of galaxy for any one of considered functions of matter distributions. Now the local value of the Newtonian "gravitation constant" measured only near the Earth. For central and peripheral parts of Galaxy a value "gravitation constant" can be vastly differ from Newtonian value. In this point of view the dark matter problem may be explain by variations of scalar field potential inside the galaxies and galaxies clusters.

*References:*

1 Eddington A.S., The Mathematical Theory of Relativity, Cambridge University Press, (1923)
2 Robertson H.P., Space Age Astronomy, ed., A.J.Deutsch and W.B.Klemperer Academic Press, (1962)
3 Will C.M., Theory and Experiment in Gravitational Physics (Cambridge University Press, Cambridge, 1981, revised 1993)
4 Bashkov V., Kozyrev S., «Problems of high energy physics and field theory», 22, Protvino, (1991).
5 Kozyrev S., Proc. 2$^{th}$ Conf. Physical Interpretation of Relativity Theory, edited by M.C.Duffy (University of Sunderland / Brit. Soc. Phil. Sci.,1992), "The problem of testing theories of gravity".
6 Jordan P., «Schwerkraft und Weltall», Braunshweig, (1955).
7 Brans C., Dicke R. Phys. Rev. 124, (1961), 925.
8 Reasenberg R.D., Shapiro I.I., MacNeil P.E., Goldsten R.B., Breidenthal J.C., Brenkle J.P., Cain D.L. and Kaufman T.M., Astrophys. J. 234, (1979), L219
9 Santiago D. I., Kallingas D., Wagoner R. V., Phys. Rev. D 56, (1997), 7627
10 Jones C., Forman W., Astrophys. J. 276, (1984), 38
11 King I., R. Astron. J. 71, (1966), 64
12 Bruckman W., F., Kazes E. Phys. Rev. D 16, (1977), 261
13 Kent, S. M. Astron. J., 93, (1987), 816
14 Battaner E., Florido E. Astro-ph/0010475 (2000)